\documentclass[epj]{svjour}
\usepackage{amsmath}
\usepackage{amssymb}
\usepackage{epsf}
\newlength{\figwidth}

\begin{document}

\title{Oscillatory behavior and enhancement of the surface plasmon linewidth in
  embedded noble metal nanoparticles}

\titlerunning{Oscillatory behavior of the surface plasmon linewidth}

\author{Rafael~A.~Molina \and Dietmar~Weinmann \and 
        Rodolfo~A.~Jalabert}

\authorrunning{R. A. Molina {\it et al.}}

\institute{Institut de Physique et Chimie des Mat\'eriaux de Strasbourg,
           UMR 7504, CNRS-ULP\\ 23 rue du Loess, 67037 Strasbourg
           Cedex, France}

\date{September 6, 2002}

\abstract{
 We study the Landau damping of the surface plasmon resonance
 of metallic nanoparticles embedded in different
 environments of experimental relevance. Important oscillations of
 the plasmon linewidth as a function of the radius of the
 nanoparticles are obtained from numerical calculations based on the
 time dependent local density approximation. These size-oscillations
 are understood, within a semiclassical approximation, as a
 consequence of correlations in the spectral density of the
 nanoparticles. We treat inert matrices, as well as the case with
 an unoccupied conduction band. In the latter case, the plasmon
 lifetime is greatly reduced with respect to the inert case,
 but the non-monotonous size-dependence persists.
}
\maketitle


\section{Introduction}
\label{sec:intro}

In recent years, the use of femtosecond 
laser pulses has made it possible to investigate the time
evolution of collective and quasi-particle excitations in metallic
nanoparticles. In these experiments, the 
actual time-resolved evolution of the energy transfer to the environment 
can be studied \cite{Bigot1,Bigot2,Feldmann}.
A determining factor in the relaxation processes is the lifetime of
the collective excitations (surface plasmons), which for small cluster
sizes ($0.5 \ nm \lesssim R \lesssim 2\,{\rm nm}$), is limited by
Landau damping (decay into electron-hole pairs). The lifetime of the
surface plasmon resonance is thus, not only a problem of fundamental
interest, but also important for possible applications.

In the last decade, the analogy of surface plasmons 
with nuclear giant resonances has been
exploited, in order to obtain a physical
understanding, as well as analytical and numerical estimations, of the
plasmon lifetime \cite{barma,yann_broglia,Yann93,MWJ}. Following these
approaches, and using a semiclassical formalism, an oscillatory
behavior of the level width was determined for free clusters of
small sizes (a number of atoms $N$ between $N=20$ and $N=1000$).
These analytical results have been verified against numerical
calculations within the time-dependent local density (TDLDA) for
a jellium model \cite{Ekardt}, and the oscillations in the level
width with the size of the nanoparticles were shown to arise from the
electron-hole density-density correlations in the angular restricted
density of states.

The non-monotonous size-dependence of the plasmon lifetime has been
experimentally observed in free alkaline metal nanoparticles
\cite{brechignac,mochizuki}, as well as in embedded noble metal
clusters \cite{charle}. Further experimental and theoretical work
would be needed in order to unambiguously characterize the level width
oscillations. While the analytical calculations are most easily
done for free alkaline clusters, metallic noble metal clusters 
embedded in a transparent matrix are more conveniently
investigated from the experimental point of view \cite{Bigot2}. 
It is then important to study theoretically the size-dependence 
of plasmon lifetime for embedded particles.
This is the purpose of this work, where
we extend the results of Ref. \cite{MWJ} to different kinds of
environments. 

The problem of the plasmon lifetime in metallic nano\-particles embedded
in a matrix, is considerably more difficult, and less understood, than
in the case of free clusters. The most obvious effect when the
clusters are embedded in a dielectric material, is the change produced by the
dielectric constant of the matrix, that moves the position of the
plasmon resonance according to the Mie formula \cite{book_K}, and can
also affect the width of the plasmon \cite{MWJ}. Moreover, when the
matrix is not inert, it can react with the atoms in the surface of the
nanoparticle. The electronic states on the surface are modified and contribute
to the loss of coherence of the collective state. 
This is the so-called chemical interface damping. 
In addition, the presence of a conduction
band in the matrix at an energy accessible to the electrons,
weakens the confinement, and
contributes to the width of the surface plasmon in an important
way. All these different effects can add up and make the theoretical
description of the system quite involved.

The experiments of Charl\'e {\em et al.} \cite{charle}
studied the influence of the matrix environments on surface plasmon 
excitations of small silver particles. They found a strong broadening 
of the surface plasmon resonance when the particles were embedded 
in a reactive matrix such as CO, as compared to an inert matrix 
from a noble gas like Ar. This effect has been attributed to
chemical interface damping. Also, a very strong broadening has been 
observed by H\"ovel {\em et al.} \cite{kreibig} for silver particles
embedded in SiO$_2$. 
In a very interesting and pioneering paper,
Persson proposed a model for explaining these experimental observations that
took into account the modification of the 
electronic eigenstates of the metallic nanoparticle by the layer 
of absorbates on its surface and by a conduction band \cite{Persson}. 

In the next section we recall the basic theoretical facts of the
calculation of the plasmon lifetime within a semiclassical
approximation. We then study the linewidth of embedded clusters from
TDLDA calculations for inert matrices. The case of matrices with 
a conduction band is treated in Sec. \ref{sec:sio2}. The
theoretical treatment of the chemical interface damping is out of
reach of our theoretical description, but the cases that we will
consider are not thought to be considerably influenced by this
effect.

\section{Theoretical background for free clusters}
\label{sec:LD}

The Landau damping of the dipole plasmon can be calculated treating
the collective excitation as an external perturbation, which can give
rise to the creation of electron-hole excitations \cite{book_BB}. 
Then, Fermi's golden rule yields the linewidth
\begin{equation}\label{Fermi}
\Delta \Gamma (R)= 2\pi \sum_{ph} \left|\langle p | \delta V | h \rangle
\right|^2 \, \delta(\hbar\omega_{\rm M} - \epsilon_p + \epsilon_h)\, ,
\end{equation}
where $| p \rangle$ and $| h \rangle$ are electron and hole states in the
self-consistent field, with its energies given by $\epsilon_p$ 
and $\epsilon_h$, respectively, $\delta V$ is the dipole field due
to the surface plasmon, and $\omega_{\rm M}$ is the frequency of the plasmon,
which classically is given by the Mie formula,
\begin{equation}\label{Mie}
\omega_{\rm M}=\frac{\omega_{\rm p}}{\sqrt{\epsilon_d+2\epsilon_m}},
\end{equation} 
where $\omega_{\rm p}$ is the plasma frequency, $\epsilon_d$ is the dielectric
function of the $d$ electrons and $\epsilon_m$ is the dielectric function
of the embedding matrix

In the case of spherical symmetry,
assuming that the confinement and the interactions lead to hard walls
at radius $R$ in the self-consistent field, we can evaluate 
Eq. (\ref{Fermi}). Integrating over the
electron-hole states, one obtains \cite{yann_broglia}
\begin{eqnarray}\label{gamma_density}
&&\Delta \Gamma (R) = c \gamma \!\!\! \int\limits_{E_{\rm F}}^{E_{\rm F}+
   \hbar\omega_{\rm M}} \!\!\!{\rm d}E\, 
  \sum_L \sum_{L'=L\pm 1} (2L+1)(2L'+1) 
\\ \nonumber 
&&\times 
<L,0;1,0|L',0>^2 
E (E-\hbar\omega_{\rm M})d_L(E)d_{L'}(E-\hbar\omega_{\rm M})\, ,  
\end{eqnarray}
where $<L,0;1,0|L',0>$ is a Clebsch-Gordan coefficient, 
$\gamma=(2\pi\hbar^3)/(3NM^2\omega_{\rm M}R^4)$, $c=4MR^2/\hbar^2$, 
$E_{\rm F}$ is the Fermi energy, and $d_L(E)$ is the one-dimensional 
density of states with total angular momentum $L$. 

Using the semiclassical expression for the density of states
of the one-dimensional problem
\cite{gutz_book}, we can decompose $d_L(E)$ in its smooth (zero-length
trajectories) and
oscillating components (arising from periodic orbits). 
We then obtain two contributions to
the width of the plasmon resonance \cite{MWJ} 
\begin{equation}\label{Gammatwo}
\Delta \Gamma=\Gamma_0+\Gamma_{\rm osc}.
\end{equation}
The smooth term $\Gamma_0$ arises from the
smooth component of $d_{\rm L}$ and exhibits the well known $1/R$ dependence 
firstly proposed by Kawabata and Kubo\cite{Kubo,yann_broglia}   
\begin{equation}\label{gamma_0}
\Gamma_0(R)=\frac{3\hbar}{4}\frac{v_{\rm F}}{R} g(\xi)\, ,
\end{equation}
where $\xi=\hbar \omega_{\rm M}/E_{\rm F}$ , $g(\xi)$ is a smoothly 
decreasing function with $g(0)=1$, and  $v_{\rm F}$ is the 
Fermi velocity.

The oscillating part  $\Gamma_{\rm osc}$ arises from the
density oscillations as a function of the energy. 
Within a semiclassical approach, it can be written as \cite{MWJ}
\begin{equation}\label{Gammaosc}
\Gamma_{\rm osc} \approx \frac {6 \sqrt{2\pi} \hbar}{M R^{2} \sqrt{k_{\rm F}
R \xi^{3}}} \sum_{r=1}^{\infty} \frac{1}{\sqrt{r}} \,
\cos(2rk_{\rm F} R \xi),
\end{equation}
where the sum runs over all repetitions $r$ of the period
of the equivalent one-dimensional motion.
The amplitude of this oscillations can be of the order of 
$\Gamma_0$ for small clusters.
\section{Embedded metallic clusters in inert matrices}
\label{sec:embedded}

\begin{figure}[tb]
\centerline{\epsfxsize=\figwidth\epsffile{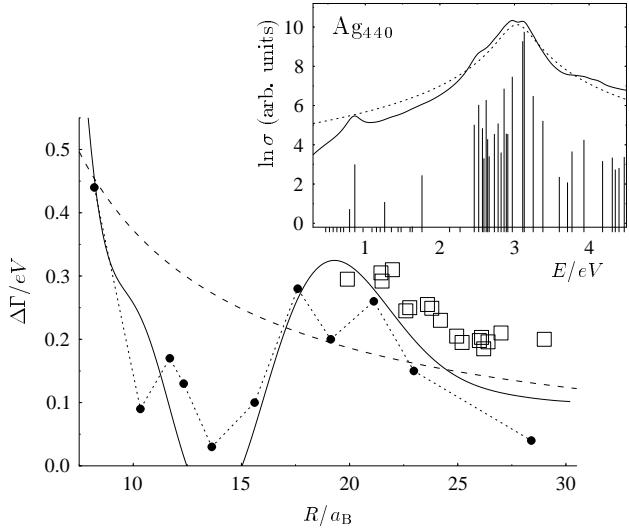}}
\vspace{3mm}
\caption[]{\label{ag} 
  Linewidth as a function of the radius (in units of the Bohr radius
  $a_{\rm B} = 0.53\, {\rm \AA}$) for silver nanoparticles in an Ar matrix
  calculated within TDLDA (full circles) together with the
  experimental results of Ref. \cite{charle} (empty squares). The
  dotted line through the numerical points is a guide-to-the-eye.
  The dashed and the solid lines represent, respectively, $\Gamma_0(R)$ and 
  $\Delta\Gamma(R)$ according to Eqs. (\ref{Gammatwo}),
  (\ref{gamma_0}) and (\ref{Gammaosc}) (with a reduction factor of 3
  as discussed in the text). Inset: logarithm of the TDLDA
  absorption cross section for Ag$_{440}$ ($R/a_{\rm B}=2.2$), showing the
  pronounced surface plasmon resonance, fitted by a Lorentzian (dotted
  line). The excited states are indicated by tick marks and their
  oscillation strengths given by the height of the vertical lines.}
\end{figure}
We want to address the question of what happens with the oscillations 
and the typical value of the linewidth of 
the plasmon when the nanoparticles are embedded in a matrix. In Ref.
\cite{MWJ} we presented some calculations for noble metals embedded in
inert matrices. The numerical results were obtained using the TDLDA
formulation by Bertsch \cite{computer} but modifying the residual
interaction to include the frequency-dependent dielectric function
of the $d$ electrons $\epsilon_d(\omega)$ and the dielectric constant of
the matrix $\epsilon_m$ \cite{Rubio}. 
In order to maintain the spherical symmetry of our problem
we always considered cluster sizes corresponding to magic numbers of
atoms.

\begin{figure}[tb]
\centerline{\epsfxsize=\figwidth\epsffile{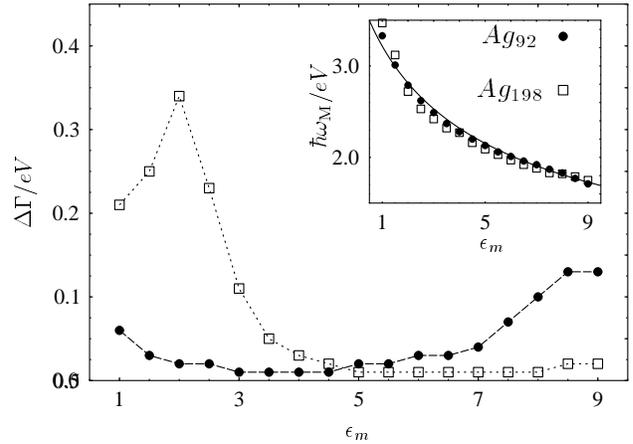}}
\vspace{3mm}
\caption[]{\label{em} Width of the surface plasmon resonance for
  silver clusters of two different sizes as a function of
  $\epsilon_m$. Inset: frequency of the plasmon $\omega_M$
  together with the expression for the Mie formula (solid line)
  with a small modification of the bulk $\omega_{\rm p}$.}
\end{figure} 
For inert matrices, the plasmon linewidth is affected only by the
modification of the frequency $\omega_{\rm M}$ through the
parameter $\xi$ of Eqs. (\ref{gamma_0}) and (\ref{Gammaosc}).
This is due to the fact that the self-consistent potential, and,
therefore, the electron-hole density-density correlation function
are practically unchanged by $\epsilon_m$, but the energy position
of the plasmon peak follows the Mie formula. 
Results for the width of the plasmon in the case of Ag
nanoparticles embedded in a matrix of Ar are shown in Fig. \ref{ag}.
In the inset we show a typical spectrum of the photo-absorption
cross-section for Ag$_{440}$ embedded in Ar.
The singularities of the spectrum are smeared out by a non-zero
$\gamma$. This value of $\gamma$ is subtracted at the
end of the calculation and we verify that the fit 
of $\Delta\Gamma$ is not sensitive
to it, provided that it is larger than a minimum value. 
$\Delta\Gamma$ exhibits
pronounced oscillations as a function of the radius. The smooth part
of the linewidth, and our semiclassical result, are an overall factor of
3 larger than the numerical results. Therefore, they have been
rescaled for in the figure. 
The experimental values of Charl\'e {\it et al.} (empty squares)
are relatively well described by the TDLDA results (within a 20 \%).
The oscillations are suppressed due to the large particle sizes and
the smearing resulting from wide size-distributions. The difference
between the Kubo formula, on one hand, 
and the numerical and experimental results, on the other hand, is
not understood. In any case, given the approximations used in the
semiclassical calculation, it is not surprising that, although, this
simple theory works
well with alkaline metals, when we include $\epsilon_d(\omega)$ there
are factors not taken into account. 
Nevertheless, the analytically obtained 
oscillations have the correct period and relative size.

Moving through different values of  $\epsilon_m$ 
we change the position of the plasmon frequency $\omega_{\rm M}$. 
Since the electron-hole density-density correlation function is
insensitive to $\epsilon_m$, the plasmon lifetime
behaves in a non-monotonous fashion as a function of $\epsilon_m$. Thus, we can
use the dielectric constant of the embedding medium as a probe for this
density-density correlation function. As an example, we show in 
Fig. \ref{em} the results of TDLDA calculations for the
width of the surface plasmon resonance for silver clusters of two
different sizes ($N=92$ and $N=198$) as a function of $\epsilon_m$.
The oscillations in the width are very different for the two
sizes due to the shell filling and the corresponding density
correlation function. Although the range of $\epsilon_m$ over which we
can see the complete oscillation of the plasmon lifetime is very large,
it should be possible to observe maxima and minima in experiments,
if we are able to change the 
dielectric constant of the surrounding medium. In the
inset of this figure we show the position of the plasmon peak as
a function of the $\epsilon_m$ for the corresponding  sizes. The
behavior with $\epsilon_m$ is very well described by the Mie formula,
and the two sizes behave in the same way. We can see almost
no difference in the position of the plasmon peak between the sizes
for the same value of $\epsilon_m$, but there are great changes in its width.

\section{Ag nanoparticles embedded in a matrix with a conduction band}
\label{sec:sio2}

\begin{figure}[tb]
\centerline{\epsfxsize=\figwidth\epsffile{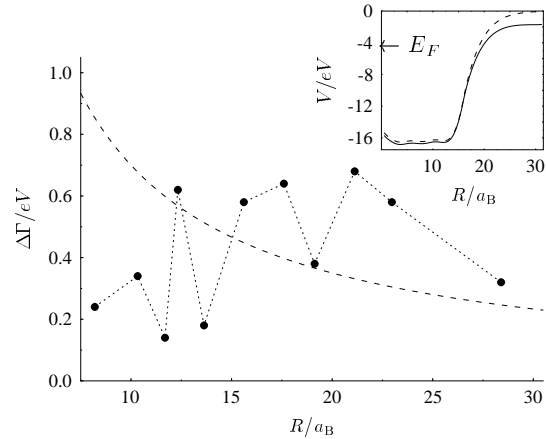}}
\vspace{3mm}
\caption[]{\label{sio} Width of the surface plasmon resonance 
  as a function of the radius of the particle for
  silver clusters embedded in SiO$_2$ (the matrix is assumed to
  support a conduction band, see text). 
  The discontinuous line is the best fit to $1/R$
  behavior. Inset: comparison of the 
  self-consistent potential $V(r)$
  with (full line) and without (dashed line) the conduction band. We
  indicate the position of the Fermi energy with an arrow.}
\end{figure}

The SiO$_2$ used in experiments for embedding Ag nanoparticles is an
amorphous solid with a conduction band with a minimum situated at 
$-1.7 \, {\rm eV}$
with respect to the vacuum energy. The valence band maximum occurs 
$10.6 \, {\rm eV}$ below the vacuum energy and has no influence in the
Ag-surface plasmon \cite{Persson}. 
Chemical interaction in the surface between Ag and
the SiO$_2$ is not expected to occur \cite{kreibig} and chemical
interface damping does not influence the width of the plasmon in this
case. In order to implement the
TDLDA calculations, we simulate the embedded medium by a change in the
boundary conditions for the calculation in the 
self-consistent potential $V(r)$ so
that $V(r) \rightarrow -1.7\, {\rm eV}$ when $r \rightarrow \infty$.
In this way the electrons are less bounded to the cluster and can go
to the conduction band in the matrix. 
The results of the self-consistent calculation of $V(r)$ and
its comparison with the potential without conduction band are shown in
the inset of
Fig. \ref{sio}. We can clearly see that the electrons at
$E_{\rm  F}$ are less tight, which translates in a small redshift
in the position of the plasmon peak due to the bigger spill-out
and into an increase (by a factor of 2) of the width 
(compare with Fig. \ref{ag}). 
This is consistent with the factor of 3
enhancement observed in the experiments of Ref. \cite{kreibig}.
Although in these experiments the distribution of sizes 
has certain unknown variance that can increase the results for the linewidth
of the optical absorption experiments, a more refined model 
including the self-interaction correction to the TDLDA, and a 
corrected dielectric function for the surface, are probably
needed for a more quantitative agreement.
However, we can clearly see that the non-monotonous behavior is
maintained and is of the same order as for free Ag
nanoparticles. Other different kinds of glasses that are used in
experiments should produce similar increases in the width of the plasmon
if the difference between the energy of the minimum of the conduction band 
and $E_{\rm F}$ is not too far from the energy of the plasmon.

\section{Conclusions}
\label{sec:conclusions}

As other size-dependent phenomena in clusters, the width of the plasmon
presents a leading-order (smooth) contribution (that goes like $1/R$), 
with oscillatory corrections due to shell effects.
For the plasmon lifetime these size-dependent corrections are much
more important than for the position of the resonance. Going for
free to embedded clusters makes this difference still more pronounced.

We have shown in this work that the size-dependent oscillations of the
linewidth, in free as well as in embedded clusters, arise from electron-hole
density correlations. Such an effect can be explored by
varying the dielectric constant of the matrix
around the nanoparticle. 

We have also considered the width of the plasmon
for a simple model of nanoparticles of Ag surrounded by a matrix of 
amorphous SiO$_2$ with a conduction band. The width is increased in a
way that agrees semi-quantitatively with the experiments. 
The oscillations in the
width are still present and could be seen in experiments with a narrow
distribution of sizes. An adequate choice of the
composition of the matrix should make it possible to increase the
lifetime of the surface plasmon by tuning its energy to a value where
the electron-hole correlation function is nearly zero, which can be 
useful in certain applications. 

\vline

We thank J.-Y.\ Bigot, H.\ Haberland, and P.A.\ Hervieux for 
helpful discussions.
RAM acknowledges the financial support from the European Union's
Human Potential Program (contract HPRN-CT-2000-00144).


\begin{thebibliography}{99}

\bibitem{Bigot1}
  J.-Y.\ Bigot, J.-C.\ Merle, O.\ Cr\'egut, A.\ Daunois,
  Phys.\ Rev.\ Lett.\ {\bf 75}, 4702 (1995).

\bibitem{Bigot2}
  V. Halt\'e, J. Guille, J.-C.\ Merle, J. I. Perakis, J.-Y.\ Bigot,
  Phys.\ Rev.\ B {\bf 60}, 11738 (1999);
  J.-Y.\ Bigot, V.\ Halt\'e J.-C.\ Merle, A.\ Daunois,
  Chem.\ Phys.\ {\bf 251}, 181 (2000).

\bibitem{Feldmann}
  T. Klar, M. Perner, S. Grosse, G. von Plessen, W. Spirkl,
  J. Feldmann, Phys.\ Rev.\ Lett.\ {\bf 80}, 4249 (1998).

\bibitem{barma}
  M.\ Barma, V. Subrahmanyam,
  J. Phys.: Condens. Matter {\bf 1}, 7681 (1989).

\bibitem{yann_broglia}
  C.\ Yannouleas and R.A.\ Broglia,
  Ann.\ Phys.\ (N.Y.) {\bf 217}, 105 (1992).

\bibitem{Yann93}
  C.\ Yannouleas, E.\ Vigezzi, R.A.\ Broglia,
  Phys.\ Rev.\ B {\bf 47}, 9849 (1993);
  C.\ Yannouleas,
  Phys.\ Rev.\ B {\bf 58}, 6748 (1998).

\bibitem{MWJ}
  R.A.\ Molina, D.\ Weinmann and R.A.\ Jalabert,
  Phys.\ Rev.\ B {\bf 65}, 155427 (2002).

\bibitem{Ekardt}
  W.\ Ekardt,
  Phys.\ Rev.\ B {\bf 31}, 6360 (1985).

\bibitem{brechignac}
  C. Br\'echignac, Ph. Cahuzac, J. Leygnier, A. Sarfati,
  Z. Phys. D {\bf 12}, 2036 (1993).

\bibitem{mochizuki}
  S. Mochizuki, M. Sasaki, R. Ruppin,
  J. Phys.: Condens. Matter {\bf 9}, 5801 (1997).

\bibitem{charle}
  K.P.\ Charl\'e, W.\ Schulze, B.\ Winter,
  Z.\ Phys.\ D {\bf 12}, 471 (1989).

\bibitem{book_BB} {\it Oscillations in Finite Quantum Systems}, 
                   by G.\ F.\ Bertsch and R.\ A.\ Broglia,
                    Cambridge University Press,
                    Cambridge, England (1994).

\bibitem{book_K} {\it Optical properties of Metal Clusters},
                  by U.\ Kreibig and M.\ Vollmer,
                  Springer, Berlin (1995).
    
\bibitem{kreibig}
  H.\ H\"ovel, S.\ Fritz, A.\ Hilger, U.\ Kreibig, and M.\ Vollmer,
  Phys.\ Rev.\ B {\bf 48}, 18178 (1993).

\bibitem{Persson}
  B.\ N.\ J.\ Persson,
  Surf. Sci. {\bf 281}, 153 (1993).
 

\bibitem{gutz_book} {\it Chaos in Classical and Quantum Mechanics} 
          by M.C.\ Gutwiller, Springer-Verlag, Berlin (1990).

\bibitem{Kubo}
  A.\ Kawabata, R.\ Kubo
  J.\ Phys.\ Soc.\ Jpn.\ {\bf 21}, 1765 (1966).

\bibitem{computer}
  G. Bertsch, Comp. Phys. Comm. {\bf 60}, 247 (1990).

\bibitem{Rubio}
  A.\ Rubio and L.\ Serra,
  Phys.\ Rev.\ B {\bf 48}, 9028 (1993).
\end{thebibliography}
\end{document}